\documentclass{article}
\usepackage{amssymb}

\title{Constructing quantum error-correcting codes for $p^m$-state
systems from classical error-correcting codes\thanks{%
To appear in IEICE Transactions on Fundamentals of Electronics,
Communications
and Computer Sciences (ISSN 0916-8508),
vol.~E83-A, no.~10, Oct.\ 2000.}}
\author{%
\begin{tabular}{c@{\hspace*{1cm}}c}
Ryutaroh Matsumoto\thanks{%
R. Matsumoto is supported by JSPS Research Fellowship
for Young Scientists.}&Tomohiko Uyematsu\\
ryutaroh@ss.titech.ac.jp&uematsu@ss.titech.ac.jp
\end{tabular}\\
Dept.\  of Electrical \& Electronic Eng.,\\
Tokyo Instutute of Technology,\\
2-12-1 Ookayama, Meguro-ku, Tokyo, 152-8552 Japan}
\date{\begin{tabular}{ll}First version:& November 4, 1999\\
Second version:& December 14, 1999\\
Third version:& June 26, 2000\end{tabular}}

\newtheorem{proposition}{Proposition}
\newtheorem{theorem}[proposition]{Theorem}
\newtheorem{lemma}[proposition]{Lemma}

\newtheorem{remark}[proposition]{Remark}
\newtheorem{corollary}[proposition]{Corollary}

\newcommand{\paper}{note}
\newcommand{\ket}[1]{|#1\rangle}
\DeclareMathAlphabet{\bm}{OML}{cmm}{b}{it}
\newcommand{\qed}{\hspace*{\fill}\rule{1.3ex}{1.3ex}}

\begin{document}
\maketitle
\begin{abstract}
We generalize the construction of quantum error-correcting
codes from $\mathbf{F}_4$-linear codes by Calderbank et~al.\ 
to $p^m$-state systems.
Then we show how to determine the error from a syndrome.
Finally we discuss a systematic construction of quantum codes
with efficient decoding algorithms.
\end{abstract}
\section{Introduction}
Quantum error-correcting codes have attracted much attention.
Among many research articles,
the most general and systematic construction is the
so called \emph{stabilizer code construction} \cite{gottesman96}
or \emph{additive code construction} \cite{calderbank97},
which constructs a quantum error-correcting code as an eigenspace
of an Abelian subgroup $S$ of the error group.
Thereafter Calderbank et~al.\ \cite{calderbank98}
proposed a construction of $S$ from
an additive code over the finite field ${\mathbf{F}}_4$ with $4$
elements.

These constructions work for tensor products of $2$-state quantum systems.
However Knill \cite{knill96a,knill96b} and Rains \cite{rains97}
observed that the construction \cite{calderbank97,gottesman96}
can be generalized to $n$-state systems
by an appropriate choice of the error basis.
Rains \cite{rains97}
also generalized the construction \cite{calderbank98}
using additive codes over ${\mathbf{F}}_4$ to $p$-state quantum systems,
but his generalization does not relate the problem of quantum code construction
to classical error-correcting codes.
We propose a construction of quantum error-correcting
codes for $p^m$-state systems
from classical error-correcting codes which is a generalization of
\cite{calderbank98}.

Throughout this \paper,
$p$ denotes a prime number and $m$ a positive integer.
This \paper\ is organized as follows.
In Section 2, we review the construction of quantum codes
for nonbinary systems.
In Section 3, we propose a construction of quantum codes for
$p$-state systems from classical codes over ${\mathbf{F}}_{p^2}$.
In Section 4, we propose a construction of quantum codes for
$p^m$-state systems from classical linear codes over ${\mathbf{F}}_{p^{2m}}$.
In Section 5, we discuss a systematic construction of
quantum codes with efficient decoding algorithms.

\section{Stabilizer coding for $p^m$-state systems}
\subsection{Code construction}
We review the generalization \cite{knill96a,knill96b,rains97}
of the construction \cite{calderbank97,gottesman96}.
First we consider $p$-state systems.
We shall construct a quantum code $Q$
encoding quantum information in $p^k$-dimensional linear space
into ${\mathbb{C}}^{p^n}$.
$Q$ is said to have minimum distance $d$ and said to be an
$[[n,k,d]]_p$ quantum code if it can detect up to
$d-1$ quantum local errors.
Let $\lambda$ be a primitive $p$-th root of unity,
$C_p, D_\lambda$ $p \times p$ unitary matrices defined by
$(C_p)_{ij} = \delta_{j-1,i \bmod p}$,
$(D_\lambda)_{ij} = \lambda^{i-1} \delta_{i,j}$.
Notice that $C_2$ and $D_{-1}$ are the Pauli spin matrices $\sigma_x$
and $\sigma_z$.
We consider the error group $E$ consisting of
$\lambda^j w_1 \otimes$ $\cdots$ $\otimes w_n$,
where $j$ is an integer, $w_i$ is $C_p^a D_\lambda^b$ with
some integers
$a,b$.

For row vectors $\bm{a} = (a_1$, \ldots, $a_n)$,
$\bm{b} = (b_1$, \ldots, $b_n)$,
$(\bm{a}|\bm{b})$ denotes the concatenated vector
$(a_1$, \ldots, $a_n$, $b_1$, \ldots, $b_n)$
as used in \cite{calderbank98}.
For vectors $(\bm{a}|\bm{b}), (\bm{a}'|\bm{b}') \in {\mathbf{F}}_p^{2n}$,
we define the alternating inner product
\begin{equation}
((\bm{a}|\bm{b}), (\bm{a}'|\bm{b}'))
= \langle \bm{a}, \bm{b}'\rangle - \langle\bm{a}', \bm{b}\rangle
\label{alternating},
\end{equation}
where $\langle,\rangle$ denotes the standard inner product in
${\mathbf{F}}_p^n$.
For $\bm{a} = (a_1$,\ldots, $a_n) \in {\mathbf{F}}_p^n$,
we define
\begin{eqnarray*}
X(\bm{a}) &=& C_p^{a_1} \otimes \cdots \otimes C_p^{a_n},\\
Z(\bm{a}) &=& D_\lambda^{a_1} \otimes \cdots \otimes D_\lambda^{a_n}.
\end{eqnarray*}
Then we have
\begin{equation}
X(\bm{a})Z(\bm{b}) X(\bm{a}')Z(\bm{b}')=
\lambda^{\langle \bm{a}, \bm{b}'\rangle -
\langle \bm{a}',\bm{b}\rangle} X(\bm{a}')Z(\bm{b}')X(\bm{a})Z(\bm{b}). \label{commute}
\end{equation}
For $(\bm{a}|\bm{b}) =
(a_1$, \ldots, $a_n,b_1$, \ldots, $b_n) \in {\mathbf{F}}_p^{2n}$,
we define the weight of $(\bm{a}|\bm{b})$ to be
\begin{equation}
\sharp \{ i \mid a_i \neq 0\mbox{ or }b_i \neq 0 \}. \label{weighta}
\end{equation}

\begin{theorem}\label{theoremp}
Let $C$ be an $(n-k)$-dimensional ${\mathbf{F}}_p$-linear subspace
of ${\mathbf{F}}_p^{2n}$ with the basis
$\{ (\bm{a}_1|\bm{b}_1)$, \ldots, $(\bm{a}_{n-k}|\bm{b}_{n-k})\}$,
$C^\perp$ the orthogonal space of $C$ with respect to the inner product
\textup{(\ref{alternating}).}
Suppose that $C \subseteq C^\perp$ and the minimum weight \textup{(\ref{weighta})}
of 
$C^\perp \setminus C$ is $d$.
Then the subgroup $S$ of $E$ generated by
$\{X(\bm{a}_1)Z(\bm{b}_1)$, \ldots, $X(\bm{a}_{n-k})Z(\bm{b}_{n-k})\}$
is Abelian, and
an eigenspace of $S$ is an $[[n,k,d]]_p$ quantum code.
\end{theorem}

Next we consider quantum codes for $p^m$-state systems,
where $m$ is a positive integer.
But the code construction for $p^m$-state systems is
almost the same as that for $p$-state systems,
because the state space of a $p^m$-state system can be regarded
as the $m$-fold tensor products of that of a $p$-state system.
We shall construct a quantum code
encoding quantum information in $p^{mk}$-dimensional linear space
into ${\mathbb{C}}^{p^{mn}}$.
For $(\bm{a}|\bm{b}) = (a_{1,1}, a_{1,2}$, \ldots, $a_{1,m}, a_{2,1}$,
\ldots, $a_{n,m}, b_{1,1}$, \ldots, $b_{n,m} ) \in {\mathbf{F}}_p^{2mn}$,
we define the weight of $(\bm{a}|\bm{b})$ to be
\begin{equation}
\sharp \{ i \mid \mbox{there exists nonzero element in}
\{a_{i,1},\ldots,a_{i,m}, b_{i,1}, \ldots, b_{i,m}\} \}. \label{weightb}
\end{equation}

\begin{corollary}\label{cora}
Let $C$ be an $(mn-mk)$-dimensional ${\mathbf{F}}_p$-linear subspace
of ${\mathbf{F}}_p^{2mn}$ with the basis
$\{ (\bm{a}_1|\bm{b}_1)$, \ldots, $(\bm{a}_{mn-mk}|\bm{b}_{mn-mk})\}$,
$C^\perp$ the orthogonal space of $C$ with respect to the inner product
\textup{(\ref{alternating}).}
Suppose that $C \subseteq C^\perp$ and the minimum weight \textup{(\ref{weightb})}
of 
$C^\perp \setminus C$ is $d$.
Then the subgroup $S$ of $E$ generated by
$\{X(\bm{a}_1)Z(\bm{b}_1)$, \ldots, $X(\bm{a}_{mn-mk})Z(\bm{b}_{mn-mk})\}$
is Abelian, and
an eigenspace of $S$ is an $[[n,k,d]]_{p^m}$ quantum code.
\end{corollary}

\subsection{Error correction procedure}\label{errorproject}
In this subsection we review
the process of correcting errors.
Let $H = {\mathbb{C}}^p$,
$H^{\otimes n} \supset Q$ the quantum code constructed by
Theorem \ref{theoremp},
and $H_{\mathrm{env}}$ the Hilbert space representing
the environment.
Suppose that we send a codeword $\ket{\psi} \in Q$,
that the state of the environment is initially
$\ket{\psi_{\mathrm{env}}} \in H_{\mathrm{env}}$,
and that we receive $\ket{\psi'} \in H^{\otimes n} \otimes H_{\mathrm{env}}$.
Then there exists a unitary operator $U$ such that
\[
\ket{\psi'} = U(\ket{\psi}\otimes \ket{\psi_{\mathrm{env}}}).
\]
If $U$ acts nontrivially $\tau$ ($0\leq \tau \leq n$)
subsystems among $n$ tensor product
space $H^{\otimes n}$, then $\tau$ is said to be
{\em the number of errors.}

We assume that $2\tau +1\leq d$,
where $d$ is as in Theorem \ref{theoremp}.
If we measure each observable in $H^{\otimes n}$
whose eigenspaces are the same as those of $X(\bm{a}_i) Z(\bm{b}_i)$
for $i=1$, \ldots, $n-k$,
where $X(\bm{a}_i)Z(\bm{b}_i)$ is as defined in Theorem \ref{theoremp},
then the entangled state $\ket{\psi'}$
is projected to $A \ket{\psi} \otimes \ket{\psi'_{\mathrm{env}}}$,
for some $A \in E$ and $\ket{\psi'_{\mathrm{env}}} \in H_{\mathrm{env}}$,
by the measurements.
By the measurement outcomes we can find
a unitary operator $A' \in E$ such that
$A'A \ket{\psi} = \ket{\psi}$.

The determination of $A'$ requires exhaustive search in general.
Thus the computational cost finding $A'$ from the measurement outcomes
is large when both $n$ and $d$ are large.
However, in certain special cases
we can efficiently determine $A'$.
An efficient method finding $A'$ is presented in Section \ref{decodep}.

\begin{remark}
The error correction method presented in this subsection
is not explicitly mentioned
in the papers \textup{\cite{calderbank97,calderbank98,gottesman96}.}
Still, it can be derived from general facts on 
quantum error correction presented in
\textup{\cite{bennett96,ekert96,knill96c}.}
A readable exposition on the error correction procedure
is provided by Preskill \textup{\cite{preskill99}.}
\end{remark}

\section{Construction of quantum codes for $p$-state systems
from classical codes}
\subsection{Codes for $p$-state systems}
In this subsection we describe how to construct quantum codes
for $p$-state systems from additive codes over
${\mathbf{F}}_{p^{2}}$.
Let $\omega$ be a primitive element in ${\mathbf{F}}_{p^2}$.
\begin{lemma}
$\{\omega, \omega^p\}$ is a basis of ${\mathbf{F}}_{p^2}$ over
${\mathbf{F}}_p$.
\end{lemma}
\noindent{\textbf{Proof:}}
When $p=2$ the assertion is obvious. We assume that $p\geq3$.
Suppose that $\omega^p = a \omega$ for some $a\in {\mathbf{F}}_p$.
Then $\omega = \omega^{p^2} = (a\omega)^p = a^2 \omega$, and
$a$ is either $1$ or $-1$.
If $a = 1$, then $\omega \in {\mathbf{F}}_p$ and $\omega$ is not a primitive
element.
If $a = -1$, then $\omega^{2p} = \omega^2$.
This is a contradiction,
because $\omega$ is a primitive element and
$2p \not\equiv 2 \pmod{p^2-1}$. \qed

For $(\bm{a}|\bm{b}) \in{\mathbf{F}}_p^{2n}$ we define
$\phi(\bm{a}|\bm{b}) = \omega \bm{a} + \omega^p \bm{b}$.
Then the weight (\ref{weighta}) of $(\bm{a}|\bm{b})$ is
equal to the Hamming weight of $\phi(\bm{a}|\bm{b})$.
For $\bm{c} = (c_1$, \ldots, $c_n), \bm{d}\in{\mathbf{F}}_{p^2}^n$,
we define the
inner product\footnote{The map
(\ref{trace}) is ${\mathbf{F}}_p$-bilinear
but does not take values in ${\mathbf{F}}_p$.
It is neither ${\mathbf{F}}_{p^2}$-bilinear
nor ${\mathbf{F}}_{p^2}$-sesquilinear.
Thus calling the map (\ref{trace}) ``inner product'' is
a little abusive.
But the map (\ref{trace}) can be converted to
an ${\mathbf{F}}_p$-bilinear form by dividing it
by $\omega^2 - \omega^{2p}$.
For this reason we call the map (\ref{trace})
``inner product''.} of $\bm{c}$ and $\bm{d}$ by
\begin{equation}
\langle \bm{c},\bm{d}^p\rangle - \langle \bm{c}^p,\bm{d} \rangle
= \langle \bm{c},\bm{d}^p\rangle - \langle \bm{c},\bm{d}^p \rangle^p\label{trace}
\end{equation}
where $\langle,\rangle$ denotes the standard inner product in ${\mathbf{F}}_{p^2}^n$
and $\bm{c}^p = (c_1^p$, \ldots,  $c_n^p)$.
For $(\bm{a}|\bm{b}),(\bm{a}'|\bm{b}') \in {\mathbf{F}}_p^{2n}$
the inner product (\ref{trace}) of $\phi(\bm{a}|\bm{b})$ and
$\phi(\bm{a}'|\bm{b}')$ is
\begin{eqnarray*}
&&\langle \phi(\bm{a}|\bm{b}), \phi(\bm{a}'|\bm{b}')^p \rangle -
\langle \phi(\bm{a}|\bm{b})^p, \phi(\bm{a}'|\bm{b}') \rangle \\
&=&
\langle \omega \bm{a} + \omega^p\bm{b} , \omega^p{\bm{a}'}^p + \omega{\bm{b}'}^p\rangle
\mbox{}- \langle \omega^p\bm{a}^p+\omega\bm{b}^p, \omega{\bm{a}'} + \omega^p{\bm{b}'}\rangle \\
&=& (\omega^2 - \omega^{2p}) (\langle \bm{a},\bm{b}'\rangle -
\langle \bm{a}', \bm{b} \rangle).
\end{eqnarray*}
Since $\omega$ is a primitive element,
$\omega^2 \neq \omega^{2p}$.
Thus the inner product (\ref{alternating}) of
$(\bm{a}|\bm{b})$ and $(\bm{a}'|\bm{b}')$ is zero iff
the inner product (\ref{trace}) of
$\phi(\bm{a}|\bm{b})$ and
$\phi(\bm{a}'|\bm{b}')$ is zero.
Thus we have
\begin{theorem}
Let $C$ be an additive subgroup of ${\mathbf{F}}_{p^2}^{n}$ containing
$p^{n-k}$ elements, $C'$ its orthogonal  space with respect to the inner product
\textup{(\ref{trace}).}
Suppose that $C' \supseteq C$ and
the minimum Hamming weight of $C' \setminus C$ is $d$.
By identifying $\phi^{-1}(C)$ with
an Abelian subgroup of $E$ via $X(\cdot)Z(\cdot)$,
any eigenspace of $\phi^{-1}(C)$ is an $[[n,k,d]]_p$ quantum code.
\end{theorem}

We next clarify the self-orthogonality of a linear code over ${\mathbf{F}}_{p^2}$
with respect to (\ref{trace}).
\begin{lemma}
Let $C$ be a linear code over ${\mathbf{F}}_{p^2}$, and
$C'$ the orthogonal space of $C$ with respect to \textup{(\ref{trace}).}
We define $C^p = \{ \bm{x}^p \mid \bm{x} \in C \}$
and $(C^p)^\perp$ the orthogonal space of $C^p$ with respect to the standard
inner product. Then we have $C' = (C^p)^\perp$.
\end{lemma}
\noindent{\textbf{Proof:}}
It is clear that $C' \supseteq (C^p)^\perp$.
Suppose that $\bm{x} \in C'$.
Then for all $\bm{y} \in C$, $\langle \bm{x}, \bm{y}^p \rangle
- \langle \bm{x}, \bm{y}^p \rangle^p = 0$.
Thus $\langle \bm{x}, \bm{y}^p \rangle \in {\mathbf{F}}_p$.
Since $\langle \bm{x}, \omega^p \bm{y}^p \rangle
- \langle \bm{x}, \omega^p \bm{y}^p \rangle^p = 0$,
$\omega^p \langle \bm{x}, \bm{y}^p \rangle \in {\mathbf{F}}_p$.
Since $\omega^p \in {\mathbf{F}}_{p^2} \setminus {\mathbf{F}}_p$,
we conclude that $\langle \bm{x}, \bm{y}^p \rangle = 0$.
\qed

\begin{theorem}\label{thm:linear}
Let $C$ be an $[n,(n-k)/2]$
linear code over ${\mathbf{F}}_{p^2}$
such that $C \subseteq (C^p)^\perp$.
Suppose that the minimum Hamming weight of
$(C^p)^\perp \setminus C$ is $d$.
Then any eigenspace of $\phi^{-1}(C)$ is an
$[[n,k,d]]_p$ quantum code.
\end{theorem}

\subsection{Error correction for $p$-state systems}\label{decodep}
In this subsection we consider how to determine the error from
measurements with
quantum codes obtained via Theorem \ref{thm:linear}.
We retain notations from Theorem \ref{thm:linear}.
Suppose that $\bm{g}_1$, \ldots, $\bm{g}_r$ is an ${\mathbf{F}}_{p^2}$-basis
of $C$.
Then ${\mathbf{F}}_p$-basis of $\phi^{-1}(C)$ is
$(\bm{a}_1|\bm{b}_1) = \phi^{-1}(\bm{g}_1),
(\bm{a}_2|\bm{b}_2) = \phi^{-1}(\omega\bm{g}_1)$, \ldots,
$(\bm{a}_{2r}|\bm{b}_{2r}) = \phi^{-1}(\omega\bm{g}_r)$.
Suppose that by the procedure in Section \ref{errorproject},
the error is converted to $A \in E$
that corresponds to $\phi^{-1}(\bm{e})$ for
some $\bm{e} \in {\mathbf{F}}_{p^2}^n$ via $X(\cdot)Z(\cdot)$,
and the original quantum state is
$\ket{\psi}$.
We know which eigenspace
of $X(\bm{a}_i)Z(\bm{b}_i)$ the vector
$A \ket{\psi}$ belongs to.
By Eq.\,(\ref{commute})
\[
X(\bm{a}_i)Z(\bm{b}_i) A \ket{\psi} = \lambda^{\ell} A\ket{\psi},
\]
where $\ell$ is the alternating inner product (\ref{alternating})
of $(\bm{a}_i|\bm{b}_i)$
and $\phi^{-1}(\bm{e})$, which is denoted by $s_i \in {\mathbf{F}}_p$.
Then we have
\begin{eqnarray*}
\langle \bm{g}_i,\bm{e}^p\rangle - \langle \bm{g}_i^p, \bm{e} \rangle &=&
(\omega^2-\omega^{2p}) s_{2i-1}, \\
\langle\omega\bm{g}_i,\bm{e}^p\rangle - 
\langle\omega^p\bm{g}_i^p, \bm{e}\rangle &=& (\omega^2-\omega^{2p}) s_{2i}.
\end{eqnarray*}
It follows that $\langle \bm{g}_i^p, \bm{e}\rangle =
(\omega^2-\omega^{2p})(\omega s_{2i-1} - s_{2i})/(\omega^p-\omega)$.
$\{ \bm{g}_1^p$, \ldots, $\bm{g}_r^p\}$ can be used as rows of the check matrix
of $(C^p)^\perp$. If we have a classical decoding algorithm for
$(C^p)^\perp$ finding the error $\bm{e}$ from a classical syndrome
$\langle \bm{g}_1^p, \bm{e}\rangle$, \ldots,
$\langle \bm{g}_r^p, \bm{e}\rangle$,
then we can find the quantum error $A \in E$.

\begin{remark}\label{omegarem}
In this section we assumed that $\omega$ is a primitive element in
${\mathbf{F}}_{p^2}$.
It is enough to assume that $\omega$ belongs to ${\mathbf{F}}_{p^2}$
and $\omega^p$, $\omega$ are linearly independent over ${\mathbf{F}}_p$.
\end{remark}

\section{Construction of quantum codes for $p^m$-state systems
from classical codes}
\subsection{Codes for $p^m$-state systems}
In this subsection we show a construction of quantum codes
for $p^m$-state systems from classical linear codes
over ${\mathbf{F}}_{p^{2m}}$.
Our construction is based on the construction
\cite{chen99} by Chen which constructs quantum codes for $2$-state systems
from linear codes over ${\mathbf{F}}_{2^{2m}}$.
We modify his construction so that we can estimate the minimum weight
(\ref{weightb}) from the original code over ${\mathbf{F}}_{p^{2m}}$.

We fix a normal basis $\{ \theta, \theta^{p}$, \ldots, $\theta^{p^{2m-1}}\}$
of ${\mathbf{F}}_{p^{2m}}$ over ${\mathbf{F}}_p$.
There always exists a normal basis of ${\mathbf{F}}_{p^{2m}}$ over ${\mathbf{F}}_p$
\cite[Section VI,\S 13]{bn:lang}.
For $\bm{a} = (a_1$, \ldots, $a_m,b_1$, \ldots, $b_m),
\bm{a}'= (a'_1$, \ldots, $a'_m,b'_1$, \ldots, $b'_m) \in {\mathbf{F}}_p^{2m}$,
we define $\phi(\bm{a}) =
a_1 \theta + a_2 \theta^p +$ $\cdots$ $+ a_m \theta^{p^{m-1}} +
b_1 \theta^{p^m} +$ $\cdots$ $+ b_m \theta^{p^{2m-1}}$,
and $T(\bm{a},\bm{a}') = 
c_{m+1} - c_1 \in {\mathbf{F}}_p$, where $\phi(\bm{a})\phi(\bm{a}')^{p^m}
= c_1 \theta +$ $\cdots$ $+ c_{2m} \theta^{p^{2m-1}}$
and $c_i \in {\mathbf{F}}_p$.
Then $T$ is a bilinear form.

\begin{lemma}
$T$ is alternating and nondegenerate.
\end{lemma}
\noindent{\textbf{Proof:}}
First we show that $T$ is alternating,
that is, $T(\bm{a},\bm{a}) = 0$ for all
$\bm{a} \in {\mathbf{F}}_p^{2m}$.
Let $x = \phi(\bm{a}) \in {\mathbf{F}}_{p^{2m}}$, and
$x x^{p^m} =
c_1 \theta +$ $\cdots$ $+ c_{2m}\theta^{p^{2m-1}}$ for $c_i \in {\mathbf{F}}_p$.
Then $(x x^{p^m})^{p^m} 
= c_{m+1} \theta + c_{m+2} \theta^p +$ $\cdots$ $+ c_{2m} \theta^{p^{m-1}} +
c_1 \theta^{p^{m}} +$
$\cdots$ $+ c_{m} \theta^{p^{2m-1}}$.
Since 
$(x x^{p^m})^{p^m} =
x^{p^m} x$,
$c_i = c_{i+m}$ for $i=1$, \ldots, $m$.
Hence $T(\bm{a},\bm{a}) = 0$.

We assume that $x \neq 0$, which implies
that $\bm{a} \neq 0$.
Since $x (\theta/x^{p^m})^{p^m} = \theta^{p^m}$,
$T(\bm{a}, \phi^{-1}(\theta/x^{p^m})) = 1$,
which shows the nondegeneracy.
\qed

\begin{lemma}
By abuse of notation,
we denote by $T$ the representation matrix of
the bilinear form $T$ with respect to the standard basis of
${\mathbf{F}}_p^{2m}$, that is,
for $\bm{a},\bm{b}\in {\mathbf{F}}_p^{2m}$,
we have $T(\bm{a},\bm{b}) = \bm{a} T \bm{b}^t$.
Let $I_{m}$ be the $m\times m$ unit matrix and
$S = \left(\begin{array}{cc}0&I_{m}\\-I_{m}&0\end{array}
\right)$.
There exists a nonsingular $2m \times 2m$ matrix $D$
such that $DTD^t = S$.
\end{lemma}
\noindent{\textbf{Proof:}} See \cite[Chapter XV]{bn:lang}
and use the previous lemma. \qed

For $\bm{c} = (c_1$, \ldots, $c_n) \in {\mathbf{F}}_{p^{2m}}^n$,
let $(a_{i,1}$, \ldots, $a_{i,m}, b_{i,1}$, \ldots, $b_{i,m}) =
\phi^{-1}(c_i) D^{-1} \in
{\mathbf{F}}_p^{2m}$.
We define $\Phi(\bm{c}) = (a_{1,1}, a_{1,2}$, \ldots, $a_{1,n}, a_{2,1}$,
\ldots, $a_{n,m}, b_{1,1}$, \ldots, $b_{n,m}) $.
Then it is clear that the Hamming weight of $\bm{c}$
is equal to the weight (\ref{weightb}) of
$\Phi(\bm{c})$, since $D$ is a nonsingular matrix.
For $\bm{a}, \bm{b} \in {\mathbf{F}}_{p^{2m}}^n$
we consider the inner product
\begin{equation}
\langle \bm{a}, \bm{b}^{p^m} \rangle, \label{innerm}
\end{equation}
where $\langle, \rangle$ denotes the standard inner product
in ${\mathbf{F}}_{p^{2m}}^n$.
\begin{proposition}\label{selfdualpm}
Let $C \subset {\mathbf{F}}_{p^{2m}}^n$ be a linear code over
${\mathbf{F}}_{p^{2m}}$,
and $C'$ the orthogonal space of $C$ with respect to \textup{(\ref{innerm}).}
Then the orthogonal space of $\Phi(C)$ with respect to \textup{(\ref{alternating})}
is $\Phi(C')$.
\end{proposition}
\noindent{\textbf{Proof:}}
For $\bm{e}=(e_1$, \ldots, $e_n), \bm{e}'=(e'_1$, \ldots, $e'_n)
\in {\mathbf{F}}_{p^{2m}}^n$,
the inner product (\ref{alternating})
of $\Phi(\bm{e}) =
(a_{1,1}$, \ldots, $a_{n,m},b_{1,1}$, \ldots, $b_{n,m})$
and $\Phi(\bm{e}') = (a'_{1,1}$,
\ldots, $a'_{n,m},b'_{1,1}$,
\ldots, $b'_{n,m})$ is
equal to
\begin{eqnarray*}
\sum_{i=1}^n \sum_{j=1}^m (a_{i,j}b'_{i,j} - a'_{i,j}b_{i,j})
&=&
\sum_{i=1}^n \phi^{-1}(e_i)D^{-1}S(D^{-1})^t\phi^{-1}(e'_i)^t \\
&=& \sum_{i=1}^n T(\phi^{-1}(e_i), \phi^{-1}(e'_i)).
\end{eqnarray*}
If $e_i e_i'^{p^m} = c_1 \theta +\cdots+ c_{2m}\theta^{p^{2m-1}}$,
then $T(\phi^{-1}(e_i), \phi^{-1}(e'_i)) = c_{m+1} - c_1$.
Thus if $\langle \bm{e}, \bm{e}'^{p^m}\rangle = 0$
then the inner product (\ref{alternating})
of $\Phi(\bm{e})$ and $\Phi(\bm{e}')$ is zero,
which implies $\Phi(C')$ is contained in the orthogonal space
of $\Phi(C)$ with respect to (\ref{alternating}).
Comparing their dimensions as ${\mathbf{F}}_p$-spaces we see
that they are equal. \qed

\begin{theorem}\label{thm:linearpm}
Let $C \subset {\mathbf{F}}_{p^{2m}}^n$ be an $[n,(n-k)/2]$ linear code over
${\mathbf{F}}_{p^{2m}}$,
$C^{p^m} = \{ \bm{x}^{p^m} \mid \bm{x} \in C \}$, and
$(C^{p^m})^\perp$ the orthogonal space of $C^{p^m}$ with respect to the
standard inner product.
Suppose that $C \subseteq (C^{p^m})^\perp$, and
the minimum Hamming weight of $(C^{p^m})^\perp \setminus
C$ is $d$.
Then the minimum weight \textup{(\ref{weightb})}
of $\Phi(C)$ is $d$,
and $\Phi(C)$ is self-orthogonal with respect to the inner product
\textup{(\ref{alternating}).}
Any eigenspace of $\Phi(C)$ is an $[[n,k,d]]_{p^m}$ quantum code.
\end{theorem}

\subsection{Error correction for $p^m$-state systems}\label{decodepm}
In this subsection we consider how to determine the error from
measurements with
quantum codes obtained via Theorem \ref{thm:linearpm}.
We retain notations from Theorem \ref{thm:linearpm}.
Suppose that $\bm{g}_1$, \ldots, $\bm{g}_r$ is an ${\mathbf{F}}_{p^{2m}}$-basis
of $C$.

Suppose that by a similar procedure to Section \ref{errorproject},
the error is converted to a unitary matrix corresponding to
$\Phi(\bm{e})$ for $\bm{e} \in {\mathbf{F}}_{p^{2m}}^n$.
We fix a basis $\{\alpha_1$, \ldots, $\alpha_{2m}\}$ of ${\mathbf{F}}_{p^{2m}}$
over ${\mathbf{F}}_p$.
Then ${\mathbf{F}}_p$-basis of $\Phi(C)$ is
$\{ \Phi(\alpha_j \bm{g}_i) \mid i=1$, \ldots, $r$, $j=1$, \ldots, $2m \}$.
First we shall show how to calculate $\langle \bm{e},\bm{g}_i^{p^m}
\rangle$ for each $i$.
For $j=1$, \ldots, $2m$,
let $(\bm{a}_j|\bm{b}_j) = \Phi(\alpha_j \bm{g}_i)$.
As in Section \ref{decodep},
by the measurement outcomes we can know the inner product (\ref{alternating})
of $\Phi(\bm{e})$ and $\Phi(\alpha_j \bm{g}_i)$, denoted by $s_j$,
for $j=1$, \ldots, $2m$.

For $x = c_1 \theta +$ $\cdots$ $+ c_{2m} \theta^{p^{2m-1}} \in {\mathbf{F}}_{p^{2m}}$, $c_1$, \ldots, $c_{2m} \in {\mathbf{F}}_p$,
we define $P(x) = c_{m+1}-c_1$.
Then $P$ is a nonzero ${\mathbf{F}}_p$-linear map.
As discussed in the proof of Proposition \ref{selfdualpm},
$s_j = P(\langle \bm{e}, \alpha_j^{p^m}\bm{g}_i^{p^m}\rangle)
= P(\alpha_j^{p^m} \langle \bm{e}, \bm{g}_i^{p^m}\rangle)$.
We define the map $P_{2m} : {\mathbf{F}}_{p^{2m}} \rightarrow {\mathbf{F}}_p^{2m}$,
$x \mapsto (P(\alpha_1^{p^m} x)$, \ldots, $P(\alpha_{2m}^{p^m} x))$.
Then $P_{2m}$ is an ${\mathbf{F}}_{p}$-linear map,
and $P_{2m}(\langle \bm{e}, \bm{g}_i^{p^m}\rangle)
= (s_1$, \ldots, $s_{2m})$.
If $P_{2m}$ is an isomorphism,
then finding $\langle \bm{e}, \bm{g}_i^{p^m}\rangle$ from
$(s_1$, \ldots, $s_{2m})$ is a trivial task,
merely a matrix multiplication.
We shall show that $P_{2m}$ is an isomorphism.

\begin{lemma}\textup{\cite[Theorem 6.1, Chapter III]{bn:lang}}
Let $W$ be a $2m$-dimensional vector space over a field $K$
with a basis $\{x_1$, \ldots, $x_{2m}\}$,
and $\widehat{W}$ the dual of $W$, that is, the $K$-linear space
consisting of linear maps from $W$ to $K$.
Then there exists a basis $\{f_1$, \ldots, $f_{2m}\}$
of $\widehat{W}$ such that $f_k(x_j) = \delta_{jk}$.
$\{f_1$, \ldots, $f_{2m}\}$ is called the \emph{dual basis}.
\end{lemma}

\begin{lemma}
There exist $\beta_1$, \ldots, $\beta_{2m}\in{\mathbf{F}}_{p^{2m}}$
such that $P(\alpha_j^{p^m} \beta_k) = \delta_{jk}$.
\end{lemma}
\noindent{\textbf{Proof:}}
Notice that $\{\alpha_1^{p^m}$, \ldots, $\alpha_{2m}^{p^m}\}$
is an ${\mathbf{F}}_p$-basis of ${\mathbf{F}}_{p^{2m}}$.
The dual space ${\widehat{{\mathbf{F}}}_{p^{2m}}}$ can be regarded as
${\mathbf{F}}_{p^{2m}}$-linear space
by defining $xf : u \mapsto f(xu)$ for $x \in {\mathbf{F}}_{p^{2m}}$
and $f \in \widehat{{\mathbf{F}}}_{p^{2m}}$.
Let $f_1$, \ldots, $f_{2m}$ be the dual basis of $\{\alpha_1^{p^m}$,
\ldots, $\alpha_{2m}^{p^m}\}$.
Since ${\widehat{{\mathbf{F}}}_{p^{2m}}}$ is one-dimensional
${{\mathbf{F}}_{p^{2m}}}$-linear space and $0\neq P \in {\widehat{{\mathbf{F}}}_{p^{2m}}}$,
$f_k$ can be written as $\beta_k P$ for some $\beta_k \in {{\mathbf{F}}_{p^{2m}}}$.
It is clear that $P(\alpha_j^{p^m} \beta_k) = \delta_{jk}$. \qed

\begin{proposition}
$P_{2m}$ is an isomorphism.
\end{proposition}
\noindent{\textbf{Proof:}}
It suffices to show that $P_{2m}$ is surjective.
For $(a_1$, \ldots, $a_{2m}) \in {\mathbf{F}}_p^{2m}$,
$P_{2m} (a_1 \beta_1 +$ $\cdots$ $+ a_{2m} \beta_{2m})
= (a_1$, \ldots, $a_{2m})$, where $\beta_k$ is as in the previous lemma.
\qed

As in Section \ref{decodep},
the error $\bm{e}$ can be determined by
a classical error-correcting algorithm for $(C^{p^m})^\perp$
from $\langle \bm{g}_1^{p^m},\bm{e}\rangle$,
\ldots, $\langle \bm{g}_r^{p^m},\bm{e}\rangle$.

\section{Notes on the construction of codes with efficient decoding algorithms}
It is desirable to have a systematic construction of quantum codes
with efficient decoding algorithms.
Calderbank et~al.\  \cite[Section V]{calderbank98}
showed a construction of cyclic linear quantum codes using the BCH bound
for the minimum distance.
With their construction we can correct errors up to the BCH bound
using the Berlekamp-Massey algorithm.

If we use the Hartmann-Tzeng bound \cite{hartmann72}
or the restricted shift bound \cite{pellikaan96}
then we get a better estimation of the minimum distance,
and we can correct more errors using modified versions of the Feng-Rao
decoding algorithm in \cite[Theorem 6.8 and Remark 6.12]{pellikaan96}.
The algorithms
\cite[Theorem 6.8 and Remark 6.12]{pellikaan96} correct errors up to the Hartmann-Tzeng bound or the restricted
shift bound.

We cannot construct good cyclic codes of arbitrary code length.
So we have to often puncture a cyclic code as in \cite[Theorem 6 b)]{calderbank98}
to get a quantum code with efficient decoding algorithms.
In classical error correction,
we correct errors of a punctured code by applying an error-and-erasure
decoding algorithm for the original code to the received word.
But there is no (classical) received word in quantum error correction.
So we decode a quantum punctured code as follows:
Let $C' \subset {\mathbf{F}}_q^n$ be a (classical) linear code,
$\bm{h}'_1$, \ldots, $\bm{h}'_r$ the rows of a check matrix for $C'$,
$C$ the punctured code of $C'$ obtained by discarding the first coordinate,
$\bm{h}_1$, \ldots, $\bm{h}_{r-1}$ the rows of a check matrix for $C$,
and $0\bm{h}_i$ the concatenation of $0$ and $\bm{h}_i$ for $i=1$,
\ldots, $r-1$.
We can express $0\bm{h}_i$ as
\[
0\bm{h}_i = \sum_{j=1}^r a_{ij} \bm{h}'_j,
\]
where $a_{ij} \in {\mathbf{F}}_q$ \cite[Lemma 10.1]{pless98}.
Suppose that an error $\bm{e} =(e_2, \ldots, e_n) \in {\mathbf{F}}_q^{n-1}$ occurs and
that we have the syndrome $s_1 = \langle \bm{h}_1, \bm{e}\rangle$,
\ldots, $s_{r-1} = \langle \bm{h}_{r-1}, \bm{e}\rangle$.
We want to find $\bm{e}$ from $s_1$, \ldots, $s_{r-1}$ using
an error-and-erasure decoding algorithm for $C'$.
We can find $s'_1$, \ldots, $s'_r \in {\mathbf{F}}_q$ such that
\[
s_i = \sum_{j=1}^r a_{ij} s'_j
\]
for $i=1$, \ldots, $r-1$.
Then there exists $\bm{e}' = (e_1,\ldots,e_n) \in {\mathbf{F}}_q^n$
such that
\begin{equation}
\langle \bm{h}'_j, \bm{e}'\rangle = s'_j \mbox{ for } j=1,\ldots, r,
\label{firstcond}
\end{equation}
because the condition (\ref{firstcond}) implies that
$\langle 0\bm{h}_i, \bm{e}' \rangle = s_i$ for $i=1$, \ldots,
$r-1$.
If we apply an error-and-erasure decoding algorithm to the syndrome
$s'_1$, \ldots, $s'_r$ with the erasure in the first coordinate,
then we find $\bm{e}'$.

Note that the algorithms \cite[Theorem 6.8 and Remark 6.12]{pellikaan96}
are error-only decoding algorithms but we can modify them to
error-and-erasure algorithms along the same line as \cite[Section VI]{shen95decomp}.

\section*{Acknowledgment}
We would like to thank Prof.\   Hao Chen,
Department of Mathematics,
Zhongshan University,
People's Republic of China,
for providing his paper \cite{chen99},
without which our paper might not be in the present form.
We would also like to thank Prof.\   Tetsuro Nishino,
Department of Information and Communication Engineering,
University of Electro-Communications,
for teaching us quantum computing.
We would also like to thank
the anonymous reviewer for the critical comments
that improved the presentation of our paper.
We would also like to thank
Dr.\  Shinji Miura,
SONY Corporation, and
Prof.\  Masaaki Homma,
Department of Mathematics,
Kanagawa University,
for pointing out Remark \ref{omegarem}.

\end{document}